# Tackling “AI against sustainability”

*Stefanie Kunkel[1,*], Malcolm Campbell-Verduyn[2], Simon Driscoll[3], Silke Niehoff[1], Gabrielle Samuel[4]*

[1] Research group “Digitalisation and Sustainability Transformations”, RIFS Research Institute for Sustainability | at the GFZ Helmholtz Centre for Geosciences Potsdam, Germany

[2] Centre for International Relations Research, Faculty of Arts, Rijksuniversiteit Groningen, Netherlands

[3] Department of Applied Mathematics and Theoretical Physics, University of Cambridge, United Kingdom

[4] Department of Global Health & Social Medicine & King’s Institute for Artificial Intelligence, King’s College London, United Kingdom

*Corresponding author; E-mail: Stefanie.kunkel@rifs-potsdam.de

## Abstract

Current debates on AI and sustainability are dichotomised, tending to focus on either the environmental impact of AI systems themselves (“sustainability of AI”) or AI’s potential for environmental benefit (“AI for sustainability”). This perspective highlights a crucial gap: “AI against sustainability” – the negative environmental consequences stemming from the application of AI technologies. While AI can offer solutions, its use in sectors like fossil fuel extraction or targeted advertising can exacerbate environmental harms, often overlooked in existing discussions. We argue for a systemic understanding of these impacts, distinguishing between AI as an object and its application, and propose a three-pronged approach to tackle “AI against sustainability” by (a) strengthened regulation, (b) proactive self-commitment by industry, and (c) constructive dialogue among stakeholders. Addressing the blind spots of “AI against sustainability” requires moving beyond isolated actions and fostering collaboration across disciplines to ensure truly more sustainable AI.

## 1. Introduction

Scientific and political debate on the relationship between environmental sustainability and AI is growing. Some researchers have characterised this debate as "polarised" [1]. Indeed, two broad sides of this debate can be identified. One side stresses the "sustainability of AI" and draws attention to the growing environmental harms of AI systems' lifecycles. This includes a focus on measuring the escalating water, material and energy consumption of AI hardware and data centres, the greenhouse gas emissions (GHGs) associated with training and deploying Large Language Models (LLMs), and e-waste [2–10]. Recommendations to reduce environmental harms of AI tend to point to the need for greater algorithmic efficiency, improved data centre energy use and cooling, as well as the integration of renewable energy sources, among others [11,12]. The other side of the debate promotes "AI for sustainability". Here, the focus is on the potential of AI to contribute to global environmental sustainability goals [13–15], for instance, via plans to use AI for enhanced climate modelling [12,16–18] or forest monitoring [19,20].

In this perspective we argue that what we identify as "AI against sustainability" remains a blind spot in bifurcated debates between "sustainability of AI" and "AI for sustainability" approaches warranting greater attention [1]. "AI against sustainability" comprises all AI applications that have intended or unintended negative environmental impacts. These impacts are insufficiently covered by "AI for sustainability" or "sustainability of AI" perspectives because these approaches tend to focus either on optimising efficiency in narrow technical senses or in searching for specific environmental use cases of AI , instead of asking how the majority of existing AI applications impact sustainability and addressing these complex sustainability challenges [15]. Therefore, we propose that the opposing impacts of AI should be viewed simultaneously and systemically instead of taking isolated actions within individual categories of AI sustainability.

To better understand and describe AI against sustainability, we first systemise the environmental impacts of AI by distinguishing positive and negative impacts of AI *as the object* and AI *as the application*. Secondly, we offer a tripartite antidote to tackle "AI against sustainability" outcomes through greater self-commitment, regulation, and constructive dialogue among stakeholders from science, industry, policymaking and civil society.

## 2. Disentangling different types of impacts of AI on the environment

Existing debates on environmental sustainability and AI tend to focus on either "sustainability of AI", which relates to the lifecycle impacts of *AI systems as objects,* or "AI for sustainability", which relates to the positive environmental impacts of *AI systems' applications*. Table 1 breaks down these positions along two axes: a) how AI systems are understood (objects or applications); b) how environmental impacts are understood as growing or shrinking. Our analysis focuses on the increases of environmental harms through *AI systems' application* (AI against sustainability) (Panel 3 in Table 1).

Table 1: Systemising different types of environmental impacts of AI systems (based on Hilty et al.[21], Kaack et al.[12], Kunkel et al.[22]).

| Direction of impact/type of impact | AI systems as objects (1st order impact) | AI systems' application (2nd and 3rd order impact) |
|---|---|---|
| **Increases in environmental harms (negative impacts)** | **(1) "Sustainability of AI"**<br>AI systems create environmental impacts along their lifecycle in their production, operation and disposal, e.g., through hardware production, energy consumption of data centres, network infrastructure etc.<br><br>Measures can be taken to enhance the "sustainability of AI", e.g., algorithmic efficiency, improved data centre energy use and cooling, and the integration of renewable energy sources. | **(2) "AI *against* sustainability"**<br>The application of AI systems creates environmental harms in an application area (2nd order), for example:<br>• AI being used in environmentally harmful sectors (e.g., fossil fuel extraction);<br>• AI being used in application areas that amplify existing environmental harms, e.g., spreading climate misinformation;<br>• systemic impacts (3rd order), e.g., when AI is being used to increase efficiency in an application area, but price changes and resulting changes in demand lead to increases rather than decreases in consumption. |
| **Reductions in environmental harms (positive impacts)** | [Because AI systems inherently consume energy and resources, reductions in environmental harms relative to a 'no-AI' baseline are unlikely.] | **(3) "AI *for* sustainability"**<br>The application of AI systems creates environmental benefits in an application area (2nd order), for example:<br>• forecasting, transparency and analysis of environmental data;<br>• optimisation of systems to reduce environmental impacts<br>• data mining and remote sensing for data-driven environmental decision making<br>• accelerated experimentation in environmental domains (e.g., clean energy) |

Environmental impacts of *AI systems as objects* (also called 1st order impacts) arise throughout the lifecycle of an AI system itself. This includes the production, operation, and disposal of physical hardware, but also the software development process, and the use of network infrastructure and data centres [12,21](panel 1 in Table 1). Technological optimisation and resulting efficiency gains in hardware, software and infrastructure are proposed in order to reduce the environmental impacts associated with AI life cycles and thereby increase the sustainability of AI [11,12].

Environmental impacts of *AI systems' applications* arise when the use of AI systems as instruments lead to changes in processes (e.g., in industry or transport) and thereby induces environmental impacts (2nd order impacts). These impacts are also sometimes called "information and communication technology-enabled impacts"[23–25]. Such environmental impacts can be positive or negative [21,26]. In any particular application domain, AI systems can create environmental benefits by facilitating transparency, monitoring and optimisation of systems, such as industrial production and transport systems (panel 3 in Figure 1). Vinuesa et al.[13] find that among the 25 environmental sub targets of the 169 Sustainable Development Goals related to climate action (SDG 13), life below water (SDG 14) and life on land (SDG 15), 93% could benefit from AI for environmental sustainability applications. Current research focuses of AI for SDGs lie on forecasting, system optimisation, data mining, remote sensing and accelerated experimentation, particularly with respect to water and vegetation[15].

However, AI systems' applications can also pose environmentally unsustainable impacts (panel 2 in Table 1). This is what we call "AI against sustainability". Productivity and efficiency gains facilitated by AI are *a priori* equally available to environmentally beneficial and environmentally damaging sectors, with the latter including, for example, fossil resource exploration, deep seabed mining, or personalised advertisement [1,12,27–29]. For instance, the World Economic Forum has reported a radical rise in innovations driven by AI that could change the mining and metals industry[30]. AI exploration is promising to extend the reach of deep-sea mining by slashing costs and boosting discovery rates[31]. Moreover, oil and gas companies have often been at the forefront of new technologies to boost exploration and production, and their adoption of AI has been no different - from reservoir simulation and remote operations to regulatory compliance and more[32]. A recent study using input-output analysis shows that in 2022, four times more investments go from the digital sector to the oil and gas sector than to the renewable and nuclear energy sector[29]. Industry analysis suggests AI and digitalisation in oil and gas can reduce upstream production costs by 10-20%[33]. By lowering extraction costs, efficiency gains through digital technologies can induce rebound effects that increase oil and gas demand, thereby slowing the phase-out of carbon-intensive activities[12,29,32].

"AI against sustainability" outcomes can also arise because of the system-wide impacts induced by the proliferation of AI, resulting from longer term changes in behaviour and economic structures (3$^{rd}$ order impacts). Importantly, AI use can lead to rebound effects that are rarely considered in existing debates. Rebound effects can arise from price changes and new production and consumption patterns that partly or fully offset energy savings promised by efficiency gains (also called "Jevon's paradox")[1,22,28,34,35]. For instance, efficiency gains in microchips and data centre energy consumption and subsequent cost reductions have led to computing power expansion and the permeation of society with AI. Furthermore, this increase in data consumption leads to (and, indeed, is hoped by many policymakers to lead to) aggregate economic growth effects[36] with related negative environmental impacts[37]. Likewise, AI enables targeted marketing that leads to higher consumption levels [38]. Consequently, growth effects resulting from AI may fully compensate for efficiency gains, leading to increasing aggregate negative environmental impacts of AI [1,34,39].

In sum, just as "AI for sustainability" can provide helpful impacts, so too can "AI against sustainability" increase environmental harms. Opposing impacts need to be viewed simultaneously and systemically, instead of isolating action within individual categories of AI sustainability.

## 3. How to mitigate "AI against sustainability"

We propose three measures for mitigating "AI against sustainability": a) regulation, (b) self-commitment, and (c) collaboration. Each of these can be enabled through better transdisciplinary collaboration of civil society, scholars, practitioners and policymakers.

### *Regulation*

It may seem like a tall order to ask for regulation that considers all impacts of AI systems' application; nevertheless, such attempts are necessary. There is a rich understanding of anticipatory governance [40] pertaining both to environmental regulation [41–43], as well as existential and catastrophic risks [44]. The oil and gas majors, as well as the energy industry as a whole, exercise forms of anticipatory governance in their scenario planning. So, too, do governments and international organizations [45]. An important step in creating 'anticipatory pressures' to regulate is making the impacts of AI applications more visible through better reporting and accounting of AI development. This could include environmental impact assessments through extended input–output analysis (EE-IOA)[29,46], and assigning legal responsibility for these impacts [47]. One tool could be stricter supply chain legislation including more stress on reporting and reducing Scope 3 up- *and downstream* environmental impacts. Upstream refers to the activities that take place prior to the companies' activities, whereas downstream refers to the activities that take place after the companies own activities. The downstream impacts would cover the application-related environmental impacts of AI systems. The aim of downstream accounting is to incentivise AI companies to divest from unsustainable businesses while creating a level playing field for all companies. An example for supply chain legislation that could cover such impacts is the European Corporate Sustainability Reporting Directive [48].

Concentrating on greenhouse gas emissions, carbon pricing promises to reduce emissions. However, the price of carbon has remained too low to incentivize sustainable practices in the context of AI. For example, the environmental impact of each individual prompt in an LLM may be near negligible whereas the aggregate energy consumption of AI data centres is expected to more than quadruple between 2024 and 2030[49]. Individual companies are unlikely to voluntarily limit the number of prompts users can make. Digital service providers could also evade transnational pricing mechanisms by moving operations to data centres in other world regions. In this context, carbon boarder adjustment mechanisms should be enhanced and extended to account for embodied environmental footprints of hardware infrastructure and digital service systems traded internationally [50].

Transnational regulations such as the EU AI Act should further recognize the importance of the application phase of AI systems. While developers or "providers" must categorise their AI as either "risky" or "not risky", and large language models must adhere to transparency measures– this is not the case for environmental harms through the application of AI [47,51]. (EU) Legislation should demand greater transparency by reporting indirect emissions, and granting public access to the reported climate information to enhance accountability and enable public scrutiny[51].

### ***Self-commitments***

A related thrust for avoiding "AI against sustainability" is encouraging AI companies and both corporate and state organisations to incorporate the impacts of AI systems' applications into their sustainability considerations. Many "Big Tech" firms already pursue sustainability commitments, for instance, proposing a lifecycle-based approach to sustainable AI design, in which environmental impact is treated as a design constraint alongside performance, cost, and reliability. Yet there are justified concerns that such commitments are eventually exposed as "greenwashing", with technology companies being unwilling to question and transform their role in proliferating unsustainable economic systems at large[52]. Moreover, there are limits to what any, even large, corporate can do on their own, so sector-level commitments seem more promising to address “AI against sustainability”.

Professional and multi-stakeholder bodies, such as the Association for Computing Machinery (ACM), the Institute of Electrical and Electronics Engineers (IEEE), and the Partnership on AI, could adopt self-commitments to account for and act on the application-related impacts of the AI systems their members develop and/or deploy. Even if self-commitments are aspirational, or more cynically, a branding and marketing exercise, such commitments could provide leverage for employees and those working within industry to enact action on avoiding “sustainability against AI” in practice[53]. Aligning commitments with related efforts to lower environmental impacts across supply chains would require building on existing initiatives rather than rethinking and redoing everything. There is a flurry of existing codes, certifications and self-imposed standards for sustainability that could be extended to include impacts of AI systems’ application[54]. For instance, the Sustainability Awareness Framework guides software developers through the thinking process of environmental and other societal impacts that their software may create. It requires companies to ask questions, such as: How does AI change my behaviour and my users’ behaviour over time and at scale, when many people use it (e.g., in mobility, consumption)[55]? Other existing tools and principles to understand and address the societal impacts of technology include Value Sensitive Design [56], Rebound Archetypes Cards [57], Responsible Research and Innovation Principles [58] and other international varieties of Technology Assessment [59]. The key message here is for interested company actors to make use of established strategies to anticipate and address technologies’ impacts across downstream users of technology.

### *Dialogues*

More inclusive dialogues amongst diverse stakeholders can further avoid “AI against sustainability”[22]. A mediating approach can support silo-breaking and encouraging interdisciplinary scholars and practitioners to engage in deeper sustainability discussions and mitigate the impacts of AI systems’ application [15,22,34,60]. Ostensibly, all stakeholders have an interest in such dialogue. For technology companies, actively avoiding the pushback emanating from growing knowledge of the application-related impacts of their products is more than a PR strategy; it can be essential in anticipating strong regulatory clampdowns should those impacts go un- or under-addressed. Creating dialogues and collaborations with interdisciplinary scholars (from human-computer interaction, sustainability science, psychology, etc.) and practitioners will also make it easier to develop effective regulation and self-commitments.

Just as with the codes and commitments above, many transnational fora already exist and can be broadened and deepened to strengthen discussions on AI’s application-related impacts. For instance, initiatives such as the Coalition for Sustainable AI, Climate Change AI, the Partnership on AI, and the Coalition for Digital Environmental Sustainability (CODES) could take up this mantle [61–64]. In addition, approaches from transdisciplinary research, such as living labs could provide spaces in which varieties of stakeholders from companies, civil society organizations, public agencies, universities, and users can collaborate in transdisciplinary manners to create, prototype, validate and test new technologies, services, products, and systems in real-life contexts [65]. There are existing attempts to develop criteria for responsible AI development in living labs[66] that can be built upon and extended to help developers and organizations quantify and act on the impacts of AI systems' application.

## 4. Conclusion

This article identified "AI against sustainability" as a missing perspective in bifurcated debates over the environmental sustainability of AI. We characterised this perspective as the outcome in which AI applications have intended or unintended negative environmental impacts insufficiently covered by "AI for sustainability" or "sustainability of AI", especially when applying AI as instruments in domains such as fossil fuel extraction, mining and marketing. Given the complex and systemic harms that "AI against sustainability" entails, we argued for avoiding these further negative outcomes via a trio of measures. These could be enacted through coalitions of industry, government, civil society and academics who together recognise the systemic nature of the challenge and seek to achieve truly more sustainable AI.

**Statement about the use of generative AI**

We used a large language model (gemma3:27b 127k provided by the GFZ Helmholtz Centre for Geosciences) for copy-editing, summarising our own text, and completing entries for the reference section.